\documentclass[a4paper,10pt]{article}
\usepackage{psfig}
\date{}
\begin{document}

\title{Competitive random sequential adsorption of point and fixed-sized particles: analytical results}
\author{{\small M. K. Hassan$^{1,2}$ and J. Kurths$^1$} \\ {\small $~^1$ University of Potsdam,
Department of Physics, Am Neuen Palais, D-14415, Potsdam, Germany} \\
 {\small $~^2$  University of Dhaka, Department of Physics, Theoretical Physics
Division, Dhaka 1000, Bangladesh}}

\maketitle

\begin{abstract}
 
\noindent
We study the kinetics of competitive random sequential adsorption (RSA) of particles of binary 
mixture of points and fixed-sized
particles within the mean-field approach. The present work 
is a generalization of the random car parking problem in the sense that
it considers the case when either a car of fixed size is parked
with probability $q$ or the parking space is partitioned into two smaller spaces 
with probability $(1-q)$ at each time event. This allows an interesting interplay
between the classical RSA problem at one extreme ($q=1$), and the kinetics of fragmentation processes 
at the other extreme ($q=0$).
We present exact analytical results for coverage for a whole range of $q$ values, 
and physical explanations are given for different aspects of the problem.
In addition, a comprehensive account of the scaling theory, emphasizing on dimensional
analysis, is presented, and the exact expression for the scaling function and exponents are obtained.
 
\end{abstract}

\vspace{3mm}
 
\noindent
PACS number(s): 05.20.Dd,02.50.-r,05.40-y
 
\vspace{2mm}

\noindent
The kinetics of adsorption or deposition of particles on a substrate is one of the most common phenomena 
that occurs in many 
branches of science and technology including physics, chemistry, biophysics and medicine. Examples include 
adsorption of macromolecules and microscopic particles such as polymers, colloid, 
bacteria, protein or latex particles \cite{kn.has1,kn.has2,kn.has3} on solid surfaces. 
Due to its importance, it has been studied extensively through all 
avenues of research comprising 
experimental, numerical and analytical means.  Owing to the complex nature of the process, one can hardly
make any progress analytically in more than one dimension and therefore most of the analytical work 
remains confined within one dimension only, where it is exactly solvable. The $1-d$ problem is not at all a 
pedagogical matter, instead it has been very useful to provide a deep understanding of the underlying 
mechanism and it has offered an insight into more realistic models in higher dimensions. It has also been 
used as a test model to verify the results obtained from numerical 
simulations and data achieved through real life experiments (for extensive review see \cite{kn.has4}).

\noindent
The simplest model one can think of that can capture the generic feature of the process is the kinetics of {\it 
random sequential adsorption} (RSA) of particles of a fixed size on a clean 
substrate \cite{kn.has14}. This is popularly known as the random car parking problem. This has 
both an instructional value as well as a long historical significance. The basic rules of the processes are 
(i) at each time event one particle is adsorbed, (ii) the sites for deposition of adsorbing particles are 
chosen randomly and (iii), once the site is chosen, the event is successful 
if the site is empty; otherwise, the particle goes back
to the particle reservoir and competes again for further pick up on an equal footing 
with the rest of the particles in the reservoir.  
Within this model, no two particles or any of their parts can occupy the same spatial position 
in the substrate. This means that particle overlapping is forbidden and hence the
resulting model describes the kinetics and structure formation of a monolayer. Moreover, once deposited, 
particles are clamped permanently, making them immobile; therefore, 
the state of the site at each time step is changed irreversibly from an empty to a 
filled state. The irreversible nature 
of the process causes the lack of a detailed balance and drives the system out of equilibrium.
Therefore, one can no longer use the well developed 
standard statistical mechanics approach fit for its equilibrium counterpart. A unique
yet fascinating feature of the RSA processes is
the existence of a jamming limit in the arbitrary dimension. 
This is the state when it is impossible to deposit even one single further particle on the substrate,
despite the presence of gaps (voids), which are in fact not large enough to allow any further deposition.

\vspace{3mm}

\noindent
To bring the RSA problem closer to real life experiment, there exist many interesting variants of this model.
Examples include cooperative sequential adsorption (CSA) \cite{kn.has4,kn.has15}, 
the accelerated random sequential adsorption (ARSA) model \cite{kn.has5}, 
the ballistic deposition (BD) \cite{kn.has6} model, the RSA on disordered substrates \cite{kn.has7},
the RSA of growing objects \cite{kn.has8} and many attempts to include transport of depositing particles by diffusion 
\cite{kn.has4}.
Recently, there has been an increasing interest in the study of the RSA 
of mixture of particles of different degree, comprising 
binary mixture of greatly differing diameters \cite{kn.has9,kn.has10}, 
and continuous mixture of sizes obeying a power-law size distribution \cite{kn.has11}.  
It is important to note that 
in the former two cases, the ultimate structure in the long time limit is described by the jamming coverage, 
whereas in the latter case, the resulting monolayer is uniquely characterised by an exponent
called fractal dimension, since the arising 
pattern is a fractal \cite{kn.has11}.
The motivation of our present work comes from Bertelt and Privman \cite{kn.has10}, 
who studied the binary mixture of monomers and
k-mers on 1-d lattice as well as Talbot and Schaaf \cite{kn.has10}, who studied the RSA of 
binary mixture of greately differing sizes. 
Bertelt and Privman obtained an approximate expression for the coverage for continuum model by taking
$k \longrightarrow \infty$. 

\vspace{3mm}

\noindent
 In the present work, we concentrate on the binary mixture of fixed size and point particles, 
instead of pointlike particles, in a $1-d$ substrate.   
We give an exact solution for the gap size distribution
function and a continuous spectrum of jamming coverage depending on the probability at which cars are parked. 
It is note worthy to mention that within the RSA model only one particle can be adsorbed
at each time step. This is the basic principle of the RSA process and it is indispensable if we want to
call it sequential adsorption.  
In our investigation we consider the case when either a car is parked with probability $q$ 
or the parking space is partitioned with probability $(1-q)$.
That is, points and
fixed sized particles compete for deposition, and at each time step only 
one of the two events is successful, and hence we call it competitive sequential adsorption. 
Indeed, we find that both the resulting dynamics and the final
coverage differ significantly from previous studies. We find that the
RSA of mixture of point-like and fixed size particles, points and fixed
size particles and binary mixture of significantly differing size
behave in a completely different way. This means that one has
to be very careful
in making any approximation, since the final results are very prone to the
exact ratio of the particle sizes.  The beauty of the present work
is that the results are exact and obtained analytically.

\vspace{3mm}

\noindent 
In addition, we present an explicit scaling description with special emphasis on the dimensional analysis. It is worth 
to mention that, despite the long history of the RSA problem, a proper description of the scaling 
theory has remained untouched, although it has been a potential candidate for this. Dimensional
analysis provides an insight into the problem and of course one cannot get things wrong if one remains
faithful to the dimensional consistency.

\vspace{3mm}
 
\noindent 
We first give the general rate equation that describes the sequential deposition of 
particles, whose sizes and positions are determined
by the specific choice of the model. Let us denote $P(x,t)$ as the gap size distribution 
function of size $x$ at time $t$, consequently 
$P(x,t)$ obeys the following integro-differential equation
\begin{equation}
{{\partial P(x,t)}\over{\partial t}} = -P(x,t)\int_0^x dz p(z) \int_0^{x-z} dy 
F(x-y-z,y|z)+2 \int_x^\infty dyP(y,t)\int_0^{y-x}dz p(z)F(x,y-x-z|z).
\end{equation}  
Here $p(z)$ is called the parking distribution function, that determines 
the size of the depositing particle and $F(x-y-z,y|z)$ 
is the deposition kernel, that determines the rate $a(x,z)=\int_0^x dz p(z)\int_0^{x-z}dy F(x-y-z,y|z)$ at which an 
interval of size $x$ is destroyed by the deposition of a particle of size $z$, thus creating two new intervals 
of size $x-y-z$ and $y$. The first term on the right hand side of
the Eq. $(1)$ describes the destruction of the interval of size $x$ and the second term 
their creation from the interval $y$ ($y>x$). Note that if one chooses $F(x,y|z)=1$
and $p(x)=\delta(x-\sigma)$ ($\sigma$ is the size of the depositing particle)  
then the Eq. ($1$) simply describes the classical car parking problem \cite{kn.has14,kn.has16}. It is 
further interesting to note that if one chooses $p(z)=\delta(z)$ then the resulting
equation describes the  random sequential deposition of points, which is equivalent to the standard binary 
fragmentation of particles \cite{kn.has13}. In fact, the car parking
problem can be interpreted as placing cuts with finite thickness and fragmenting particles into 
two pieces at each time step. The connection between the RSA
and the binary fragmentation process was first noted by Ziff \cite{kn.has12}.

\vspace{3mm}

\noindent
In order to gain a detailed insight into the process, we first study the 
RSA of point-sized particles alone, which 
in fact describes the kinetics of the
binary fragmentation process. Let us consider the simplest case when $F(x,y)=1$, which is known 
as the random scission model \cite{kn.has13,kn.charlesby};
the resulting equation is
\begin{equation}
{{\partial P(x,t)}\over{\partial t}} = -xP(x,t)+2\int_x^\infty P(y,t)dy.
\end{equation}
First, note that $a(x)=x$ is the quantity that describes the rate and therefore it must bear the dimension 
inverse of time $t$. This means that $x$ and $t$ are interlocked 
and any one of these two variables can be expressed in terms of the other one. On the other hand, since 
$x$ and $t$ are the only two governing parameters,
the quantity $P(x,t)$ can also be expressed in terms of one parameter alone. 
Therefore, assuming $t$ to be the independent parameter,
we can write the following scaling {\it ansatz}
\begin{equation}
P(x,t) \sim P_0(t) \phi(x/s(t))
\end{equation} 
where $\phi(\xi)$ is known as the scaling function. The Eq. ($2$) will admit scaling if we 
can choose a time dependent scale $s(t)$ for the spatial 
variable, and $P_0(t)$ for the particle size distribution function so that all plots 
of $P(x,t)/P_0(t)$ against $x/s(t)$ for any initial distribution
collapse onto one single curve. This can only be true if both $P_0(t)$ and 
$s(t)$ show a powerlaw behaviour. This is due to the
fact that $P_0(t)$ and $s(t)$ must bear the dimensions of $P(x,t)$ and $x$ respectively. 
On the other hand, the dimension of any physical quantity must be of 
power-monomial nature, i.e. we can write
\begin{equation}
P_0(t) \sim t^\alpha \hspace{4mm} {\rm and} \hspace{4mm} s(t) \sim t^z.
\end{equation}  
Simple dimensional analysis of Eq. ($1$) immediately reveals $z=1$ and since 
it describes the conservation of mass, i.e $\int_0^\infty x P(x,t)dx$, it is
independent of time and yields $\alpha=2$. The solution of the Eq. (1) is now well known and reads as  
\begin{equation}
P(x,t) \approx t^2 e^{-xt}
\end{equation}
which gives $\phi(\xi)=e^{-\xi}$. Therefore, both the exact solution and the dimensional analysis
confirm that $t^2$ bears the dimension of $P(x,t)$ and $x$ bears the dimension inverse of time. 
We shall use this result later.
 
\vspace{3mm}
 
\noindent 
We now study the random sequential deposition of a mixture of points and of 
particles of size $\sigma$. Considering the rate equation
for the classical car parking problem and the binary fragmentation model, we can write the following equation 
\begin{equation}  
{{\partial P(x,t)}\over{\partial t}}  =    -(x-q\sigma)P(x,t)+2q\int_{x+\sigma}^\infty dy P(y,t)   
+2(1-q)\int_x^\infty P(y,t)dy  \hspace{3mm} {\rm for}  \hspace{3mm} x \geq \sigma
\end{equation}
and
\begin{equation}
{{\partial P(x,t)}\over{\partial t}}= 2q\int_{x+\sigma}^\infty dy P(y,t) +(1-q)\Big (
 -xP(x,t)+2\int_x^\infty P(y,t)dy \Big) \hspace{3mm} {\rm for}  
\hspace{3mm} x < \sigma.
\end{equation}
This is an obvious generalization of the RSA problem, in which at each time event a particle 
of size $\sigma$ is deposited with probability $q$
or a point is deposited with probability $(1-q)$. Another way of interpreting the 
present model is that at each time event a car is parked with probability
$q$ on the one dimensional parking space or the parking space is divided 
into smaller spaces with probability $(1-q)$. 
Knowing the solution of Eq. ($2$) and the solution for the car parking problem alone,
we write the following {\it ansatz} 
\begin{equation}
P(x,t) = A(t)e^{-(x-q\sigma)B(t)}
\end{equation}
where $A(t)$ and $B(t)$ are yet to be determined.
The simplest way of finding $B(t)$ is by appreciating that the argument of an exponential 
function must be a dimensionless 
quantity. On the other hand Eq. ($6$) implies that $(x-q\sigma)$ is the rate at which gaps
of size $x$ are destroyed, and hence it must bear the dimension inverse of time. Therefore, only $B(t)\sim t$ can
make the solution physically acceptable. In order to obtain the solutions for $A(t)$ and $B(t)$ one can also substitute 
the {\it ansatz} into the Eq. (6). This gives two differential equations, (i) the differential 
for $B(t)$, ${{dB(t)}\over{dt}}=1$ 
and (ii) the differential equation for $A(t)$
\begin{equation}
{{d \ln A(t)}\over{dt}}=2q{{e^{-\sigma t}}\over{t}}+2(1-q)/t.
\end{equation}
Both the equations should be solved subject to the initial condition that we start the process with an empty substrate
i.e. $P(x,0)=0$. 
Therefore, the solution for $B(t)$ is trivial, which only confirms the solution we obtained through the dimensional
analysis. Note that the solution of the equation for $A(t)$ must be of the form $A(t)\propto t^2$, since 
$P(x,t)$ must bear the dimension of $t^2$.
It is then quite straightforward to obtain the solution of Eq. (9) which is
\begin{equation}
A(t)=t^2 F_{q} (\sigma t) 
\end{equation}
where $F_{q}(\sigma t)= e^{-2q\int_0^{\sigma t}{{(1-e^{-u})}\over{u}}du}$. 
Therefore, the solutions of Eqs. (6) and (7) are
\begin{equation}
P(x,t)=t^2 F_{q} (\sigma t)e^{-(x-q\sigma)t}  \hspace{3mm} {\rm for}  \hspace{3mm}  x \geq \sigma
\end{equation}
and 
\begin{equation}  
P(x,t)=\int_0^t d\tau \tau F_{q}(\sigma\tau) e^{-(x-q\sigma)\tau}\Big[2qe^{-\sigma \tau}+(2-\tau)(1-q)\Big] 
\hspace{3mm} {\rm for}  
\hspace{3mm}  x < \sigma.
\end{equation}
The above solutions imply that 
we can still recover the solution of the binary fragmentation process by taking $\sigma=0$.
It is important to mention that the solution of Eq. (6) alone
is enough to provide us with all the interesting information we need. 
In fact, we never use the solution for $P(x,t)$ when $x < \sigma$, which is 
also true in the case of the classical car parking problem. 

\vspace{2mm}

\noindent
To give a scaling analysis and for better clarity we define $P(x,t)\equiv P((x-q\sigma),t)$ 
for $ x \geq \sigma $, 
then we find that the solution of Eq. (6) satisfies the following identity 
\begin{equation}
P((x-q\sigma)\lambda, t/\lambda)=\lambda^{-2}P((x-q\sigma),t).
\end{equation}
This relation is the hallmark for the existence of scale invariance and it is equivalent 
to a {\it data-collapse} formalism \cite{kn.stanley}, as we
shall show below. It means that 
if the deposition rate $(x-q\sigma)$ is increased by a factor $\lambda$, and the observation
 time is decreased by the same factor,
then the resulting structure would look the same except for a numerical 
prefactor. In other words, increasing $(x-q\sigma)$ by a factor
of $\lambda$ and decreasing $t$ by the same factor means that the numerical value of the 
gap size distribution function is reduced by
a factor of $\lambda^2$. Since the Eq. (13) is true for all positive values of 
$\lambda$, we can choose $\lambda=t$ in Eq. (13) then
\begin{equation}
P_{scale} = P(\xi,1)=\phi(\xi,\eta)
\end{equation}
where
\begin{equation}
P_{scale} ={{P}\over{t^2}};   \hspace{3mm} \eta=\sigma t \hspace{3mm} {\rm and}  \hspace{3mm} \xi=(x-q\sigma)t
\end{equation}
and $\phi(\xi,\eta)$ is the scaling function. Comparing with the explicit result we get
\begin{equation}
\phi(\xi,\eta)=F_{q} (\eta)e^{-\xi}
\end{equation}
and hence $\phi$ is indeed a dimensionless quantity. Therefore,
if $P_{scale}$ is plotted against the scaled size $\xi$, then the entire family 
of curves collapses into one single
curve described by the function $\phi(\xi,\eta)$. This is possible due to the fact that the scaling 
function depends only on a dimensionless quantity,
the numerical value of which must be independent of the choice of units. 
The method of {\it data collapse} is a very powerful technique for
establishing scaling and it is especially useful to analyze and extract exponents 
from data obtained from numerical simulations or from real life
experiments.

\vspace{3mm}
 
\noindent    
The most important quantity of interest in the RSA process  
is the coverage or jamming limit which is defined as
\begin{equation}
\theta(t)=1-\int_0^\infty xP(x,t)dx.
\end{equation}
It tells us the fraction of the substrate covered by the depositing particles. In the case of 
classical random car parking problem only 74.759 percent of the total substrate is covered by the depositing particles. 
It is now interesting to see, how the deposition of points or the division of parking spaces 
changes the jamming limit. 
In the present case, of course, there is always
space for adsorption of points as the points do not have any width due to their inherent definition. 
Hence, they do not contribute to the coverage
or the jamming limit. Therefore, in this model the jamming limit is 
the state when there are no more gaps available for the adsorption of particles of size $\sigma$. 
Once this state is achieved, it is no longer necessary
to continue the process as far as the jamming limit is concerned. 
So, it would be interesting to see
in what way the coverage changes, if at all.

\vspace{3mm}

\noindent
To find this out, it is more convenient to deal with the rate 
equation for the coverage than the coverage itself. Combining Eq. (6-7) and Eq. (17) we get  
\begin{equation}
{{d\theta}\over{dt}}=q\sigma \int_\sigma^\infty dx(x-\sigma)P(x,t).
\end{equation} 
Clearly it states that only fixed size particles contribute to the coverage. 
Notice that the right hand side of Eq. (18) bears the dimension inverse of time, as it should be.
Substituting the solution of Eq. (6) into this and integrating it, 
we find the final expression for the jamming limit 
\begin{equation}
\theta(\infty)=q\int_0^\infty dsF_q(s) e^{-(1-q)s}.
\end{equation}
This is a dimensionless quantity and hence, upon transition from one unit of measurement to another 
within a given class, its numerical value must remain unchanged. Therefore, 
the final coverage or the jamming limit $\theta(\infty)$ is independent of the size of the depositing 
particle $\sigma$, which is indeed a non-trivial and interesting result. 
This also reflects the fact that the gap size distribution function $P((x-q\sigma),t)$  
satisfies the exact identity (Eq. (13)) and the function $F_{q}(\sigma t)$ is a 
dimensionless quantity. In
fact, all properties of interest are independent of the size of the depositing particle, and
the rate at which they are adsorbed is due to the existence of the scaling property.
Clearly, if $\sigma$ is increased by a factor $\lambda$, then the time to reach the 
jamming configuration is reduced by a factor $\lambda$. 
However, this is true only if the substrate is sufficiently large in comparison to the depositing particle.
One can now obtain a whole range of values for the jamming limit by just tuning 
the $q$ value. Fig. 1. shows that the jamming limit starts 
from zero at $q=0$, since there the problem reduces itself to the random 
sequential adsorption of points. Note the inherent definition
of point that does not occupy space. Therefore, one might apprehend that it 
does not play any role in achieving the jamming limit if it is adsorbed sequentially with a fixed size particle. 
However, note that once a point is deposited, it excludes fixed-sized particles from landing 
within around $\sigma /2$, which is the distance between
the point where it is chosen to land and the centre of the fixed sized particle. This exclusion will certainly 
have effect on the resulting dynamics of the process.

\begin{figure}[H]

\psfig{figure=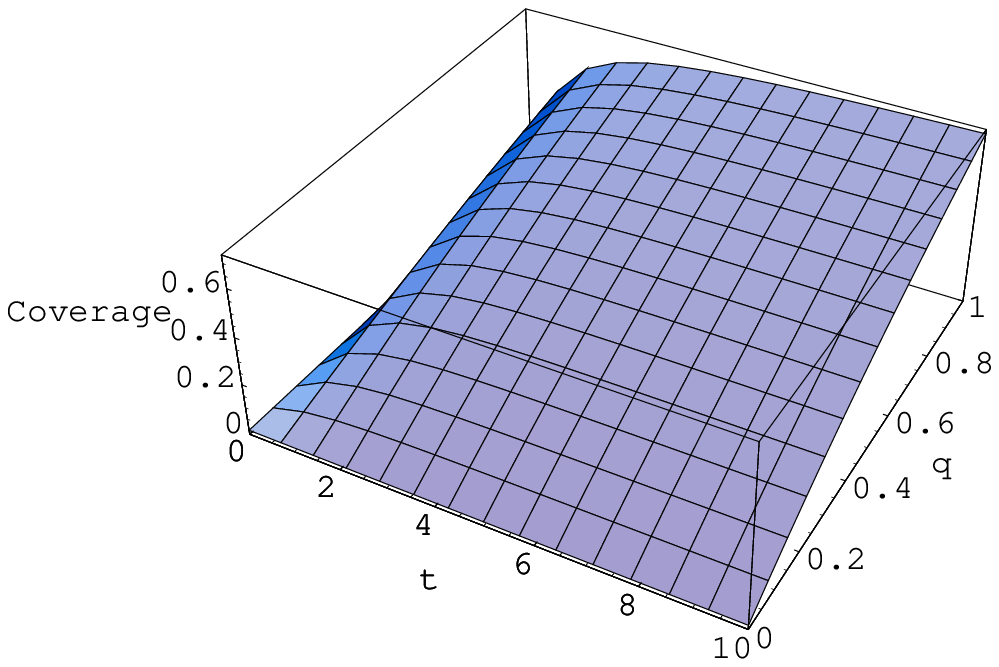,width=80mm,height=60mm,angle=0}
\caption[]{Jamming limit as a function of time $t$ and probability $q$ (a 3D view).} 
\end{figure}

\noindent
In other words, a point divides the substrate, and hence each successful 
deposition of points creates two new ends belonging 
to two different gaps. No particle can be adsorbed on a space which belongs to two different gaps i.e.
overlapping of points is forbidden.
We find that for $q>0$, the 
jamming limit changes nontrivially. Fig. (1) exhibits a $3$ dimensional view of the jamming limit
against time and the probability $q$ with which fixed-size particles are adsorbed. Fig. (2) 
is the cross-section of the coverage-time plane at $q=1$ and $q=1/2$ of fig. (1). 
It is generally believed that if a mixture contains a small number of particles of
different sizes, then the kinetics and jamming limit are primarily determined 
by the smallest size which in the present case is a point particle. However, 
in our study we find that the jamming limit is primarily determined by the larger 
particle; yet its dynamics and other aspects are influenced
by the size of the smaller particle. For $q >0$, the jamming limit is always smaller than its corresponding
classical counterpart. Fig. (2) is plotted to demonstrate the slower approach to the jamming limit 
as $q$ value decreases. 
This is due to the fact that the adsorption of points leads to the crowding of more 
partition in the parking space than to the adsorption of fixed sized particles.
Fig. (3) in fact is the cross-section  of coverage-$q$ plane of Fig. (1) at the state when the system has achieved 
the jamming limit. It shows that the jamming limmit increases monotonously as fixed-size particles
wins more in the competition with the point particles for adsorption. However, 
it does not increase linearly with $q$, instead one
can subdivide the whole curve into a number of regimes 
of width $0.2$ in $q$ values, so that in each regime they can 
be well approximated to a straight line. However, notice that, between $q=0.95$ and $q=1$,
the coverage increases much stronger than in any previous regime. The appearance of a non-linear rise
becomes more pronounced if it is plotted in the larger scale.
   
\begin{figure}[H]
\psfig{figure=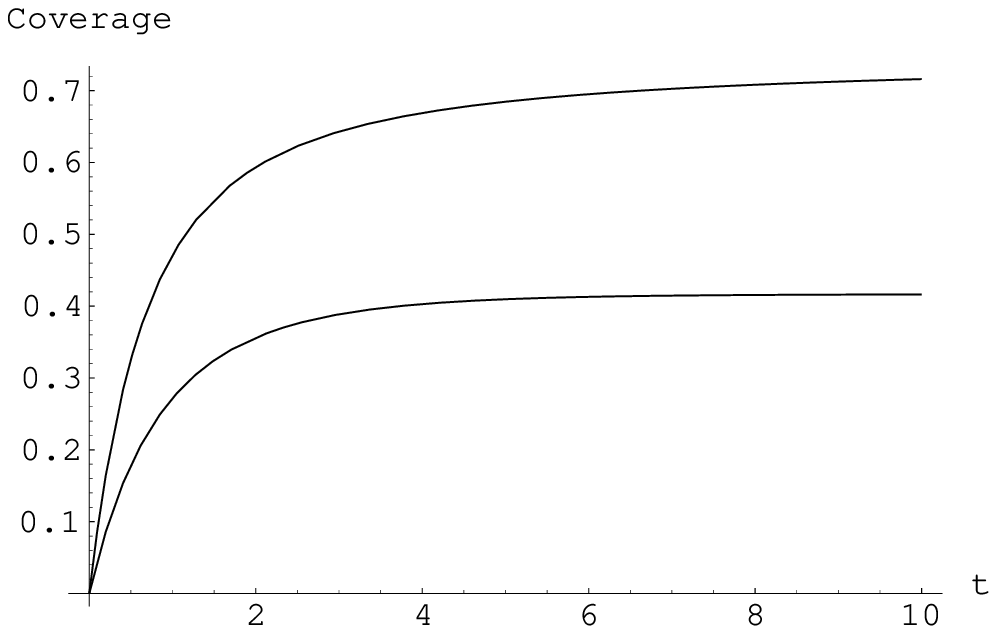,width=70mm,height=50mm,angle=0}
\caption[]{Jamming limit $\theta(\infty)$ against time $t$. The lower curve is for $q={{1}\over{2}}$ 
and the upper curve for $q=1$ (the classical car parking).}
\end{figure}

\begin{figure}[H]
\psfig{figure=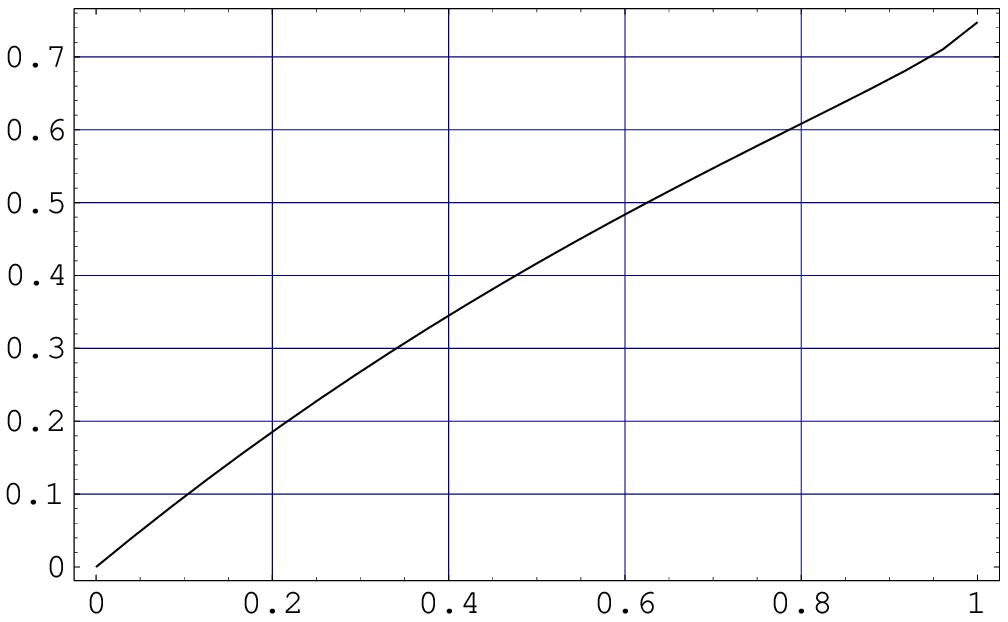,width=60mm,height=40mm,angle=0}
\caption[]{Jamming limit $\theta(\infty)$ as a function of $q$ (this is in fact $[100]$ 
face of the Fig. (1)).}
\end{figure}

\noindent
In conclusion, the present model, though simple, can yet capture some generic features of the RSA of mixture
of particles such as coverage and the scaling behaviour. We find that the coverage of mixture of points
with fixed-sized particles stays always lower than its classical counterpart. 
It would be interesting to  see how the results
change when both the contents of the binary mixture are of a finite size but differ in
length which  we intend to present in a forthcoming paper \cite{kn.preparation}. In addition, we give an extensive 
scaling description of the RSA problem which has not been addressed so far. 
One potential application of the present work could be the simple cyclization reaction process
where adjacent pendant groups link randomly along the polymer, or the polymer itself undergoes
a possible degradation. This is in fact a natural generalization of the original work by Folry
\cite{kn.has2}.
We believe
the present work will shed a new insight into the underlying mechanism of the problem and will help
guiding numerical and experimental works dealing with more realistic situations addressing the adsorption
of mixture.

\vspace{3mm}

\noindent
 We are grateful to P. L. Krapivsky and G. J. Rodgers for useful correspondence during the work. 
MKH is grateful to the Alexander von Humboldt Foundation for awarding the fellowship.

\end{document}